\documentclass[review]{elsarticle}%Physica A
\usepackage{lineno,hyperref}
\usepackage{amsmath}
\usepackage{amssymb}
\usepackage{graphicx}
\usepackage{caption}
\usepackage{subcaption}
\newtheorem{thm}{Theorem}
\newtheorem{dfn}{Definition}
\begin{document}

\begin{frontmatter}

\title{A new $\kappa$-deformed parametric model for the size distribution of wealth}

% \author{Adams Vallejos}
% \address{Radarweg 29, Amsterdam}
% \fntext[myfootnote]{Since 1880.}

\author[mymainaddress,mysecondaryaddress]{Adams Vallejos}
\author[mymainaddress,mysecondaryaddress]{Ignacio Ormaz\'abal}
\author[mymainaddress,mysecondaryaddress]{F\'elix A. Borotto}

\author[mymainaddress,mysecondaryaddress]{Hern\'an F. Astudillo\corref{mycorrespondingauthor}}
\cortext[mycorrespondingauthor]{Corresponding author}
\ead{hastudil@udec.cl}

\address[mymainaddress]{Departamento de F\'isica, Universidad de Concepci\'on, Concepci\'on, Chile}
\address[mysecondaryaddress]{Grupo de Sistemas Complejos, Universidad de Concepci\'on, Concepci\'on, Chile}

\begin{abstract}
It has been pointed out by Patriarca \emph{et al.} (2005) that the power-law tailed equilibrium distribution in heterogeneous kinetic exchange models with a distributed saving parameter can be resolved as a mixture of Gamma distributions corresponding to particular subsets of agents. Here, we propose a new four-parameter statistical distribution which is a $\kappa$-deformation of the Generalized Gamma distribution with a power-law tail, based on the deformed exponential and logarithm functions introduced by Kaniadakis(2001). We found that this new distribution is also an extension to the $\kappa$-Generalized distribution proposed by Clementi \emph{et al.} (2007), with an additional shape parameter $\nu$, and properly reproduces the whole range of the distribution of wealth in such heterogeneous kinetic exchange models. We also provide various associated statistical measures and inequality measures.
\end{abstract}


\end{frontmatter}

\section{Introduction}
During the last decades the study of income and wealth distributions has been a matter of great interest not only for economists but also for mathematicians and physicists. From a statistical physics perspective, the first agent based model\cite{Dragulescu2000} considered a closed economic system of money-exchanging agents in analogy with a particle gas of interacting particles, that is,  a large set of $N$ interacting components exchanging a (conservative) quantity $x$ representing money, income and/or wealth indistinctly. The system evolves according to a simple trading rule: in a given moment, a pair of agents $(i,j)$ with money $(x_i,x_j)$ respectively, exchange a certain amount $\Delta x$ following the rule:
\begin{subequations}
\label{KEM1}
\begin{eqnarray}
 x_i'=x_i-\Delta x,\\
 \label{KEM2}
 x_j'=x_j+\Delta x,
\end{eqnarray}
\end{subequations}
where $(x'_i,x'_j)$ stands for the remaining money of the pair after the interaction. If this procedure is iterated many times, the system approaches a steady state characterized by a distribution function $f(x)$. From (\ref{KEM1}) it is straightforward to realize that the interaction is also conservative due to the fact that $x_i'+x_j'=x_i+x_j$. Numerical results\cite{Dragulescu2000,Dragulescu2001a,Dragulescu2001b} show that, as a gas of elastically interacting particles, the distribution of money can be described by a Boltzmann-Gibbs (BG) distribution:
\begin{equation}
\label{BoltzmannGibbs}
 f(x)=\frac{1}{\langle x\rangle}\exp\left(-\frac{x}{\langle x\rangle}\right),
\end{equation}
where $\langle x\rangle$ is the average money the system. The BG distribution (\ref{BoltzmannGibbs}) is only obtained when the microscopic equations (\ref{KEM1}) have time-reversal symmetry, meaning that it is indistinguishable if money flows from $i\to j$ or from $j\to i$. The exponential distribution is highly due to the no debt condition $x_i\geq0$ imposed to the system, giving as a result an accumulation of agents with no money at all in the $x=0$ border. Thus, as the interactions take place in time, less and less agents are left to interact with each other producing a rich gets richer situation. In the particle gas analogy, the exponential distribution (\ref{BoltzmannGibbs}) is equivalent to that of kinetic energy in $D=2$ dimensions\cite{Patriarca2017a}. Nevertheless, real-world societies only match for this kind of exponential distributions in the intermediate income range\cite{Dragulescu2001a,Dragulescu2001b} and drawing away from the theoretical curve in the low and high limits. Extensions to this kind of agent based models\cite{Chakraborti2000,Chakraborti2002,Chatterjee2003,Chatterjee2004,Chatterjee2005,Chakraborti2008,Chakraborti2009} introduced a saving propensity parameter $0\leq\lambda_i<1$ into the model, representing the individual fraction of wealth the agent $i$ saves during a transaction. The trading term $\Delta x$ for this case then writes as:
\begin{equation}
\label{heterogeneous}
 \Delta x= (1-\varepsilon)(1-\lambda_i)x_i-\varepsilon(1-\lambda_j)x_j,
\end{equation}
where $\varepsilon$ is a random number uniformly distributed between 0 and 1 and $\lambda_i$ is the saving propensity of the $i$-th agent. In the first place, the homogeneous version where all the agents have the same saving propensity $\lambda_i\equiv\lambda$ drives the system into a Gamma-like steady state distribution\cite{Patriarca2004}:
\begin{equation}
\label{gamma}
 f(x)=\frac{n}{\langle x\rangle}~\frac{1}{\Gamma(n)}\left(\frac{n}{\langle x\rangle}x\right)^{n-1}\exp\left(-\frac{n}{\langle x\rangle}x\right),
\end{equation}
with $n=n(\lambda)$ is given by:
\begin{equation}
 n(\lambda)=\frac{3\lambda}{1-\lambda}+1,
\end{equation}
where $0\leq n<\infty$. This is consistent with results proposed in the literature\cite{SalemMount1974, Angle1986,Angle1993,Angle2002}. Specifically, by setting $\lambda=0$ the equilibrium distribution (\ref{gamma}) with $n=1$ reduces to the exponential distribution (\ref{BoltzmannGibbs}), and again, in the kinetic theory perspective, the distribution (\ref{gamma}) represents the canonical Boltzmann equilibrium distribution for a gas in $D=2n$ dimensions\cite{Patriarca2017a}. Secondly, if the $\lambda_i$'s in (\ref{heterogeneous}) are not all equal, the wealth distribution of the system shows exponential form in the mid-income range and a Pareto power-law form in the high income range\cite{Pareto1896}. It is shown that the overall distribution corresponds to a statistical mixture of subsystems of $\lambda_i$, which are individually Gamma distributed\cite{Patriarca2005,Patriarca2017b} as described by (\ref{gamma}). Although the microscopic origin of power laws is still undetermined, it is found that in many complex systems this kind of heavy tailed distributions emerge naturally due to the introduction of heterogeneity or diversity into the system. Empirical observations\cite{Pareto1896} state that the wealthier group distribution follows a power-law behaviour $f(x)\sim x^{-a-1}$, where the Pareto exponent $a$ varies from 1 to 2. Kinetic exchange models of economy\cite{Das2003,Repetowicz2005,Chatterjee2005} show that this Pareto exponent for a steady state distribution of wealth is equal to one for the wealthier group. The present paper introduces a four-parameter distribution which is an extension to the $\kappa$-Generalized ($\kappa$G) distribution proposed by Clementi \emph{et al.}\cite{Clementi}. This new distribution has four parameters and is a $\kappa$-deformation of the Generalized Gamma distribution based on the $\kappa$-exponential and $\kappa$-logarithm functions proposed by Kaniadakis\cite{Kaniadakis2001,Kaniadakis2002,Kaniadakis2005}:
\begin{subequations}
\label{deformedfunctions}
\begin{eqnarray}
 \exp_{\{\kappa\}}(x)=\left(\sqrt{1+\kappa^2x^2}+\kappa x\right)^{\frac{1}{\kappa}},~x\in\mathbb{R}\\
 \ln_{\{\kappa\}}(x)=\frac{x^\kappa-x^{-\kappa}}{2\kappa},~x\in\mathbb{R_+}
\end{eqnarray}
\end{subequations}
which are limiting cases of both the ordinary exponential and logarithm functions as $\kappa\to0$. The previous $\kappa$-functions show interesting properties for the analysis of income and wealth distributions due to their power-law asymptotic behaviour:
\begin{subequations}
\begin{equation*}
 \lim_{x\to\pm\infty}\exp_{\{\kappa\}}(x)\sim |2\kappa x|^{\pm\frac{1}{|\kappa|}},
\end{equation*}
\begin{equation*}
 \lim_{x\to0^+}\ln_{\{\kappa\}}(x)=-\frac{1}{|2\kappa|}x^{-|\kappa|},
\end{equation*}
\begin{equation*}
 \lim_{x\to+\infty}\ln_{\{\kappa\}}(x)=\frac{1}{|2\kappa|}x^{|\kappa|}.
\end{equation*}
\end{subequations}
In the framework of $\kappa$-deformations\cite{Kaniadakis2013} some special functions can be obtained. In particular, let $x$ be any complex variable, the generalized Gamma function $\Gamma_{\{\kappa\}}(x)$ is defined as:
\begin{equation}
 \Gamma_{\{\kappa\}}(x)=[1-\kappa^2(x-1)^2](x-1)\int_0^\infty t^{x-2} \exp_{\{\kappa\}}(-t)~dt,
\end{equation}
with $\Gamma_{\{\kappa\}}(1)=\Gamma_{\{\kappa\}}(2)=1$ and $\Gamma_{\{\kappa\}}(3)=2$, and terms of the standard $\Gamma(x)$ functions, is given by:
\begin{equation}
\label{kgammaf}
 \Gamma_{\{\kappa\}}(x)=\frac{1-|\kappa|(x-1)}{|2\kappa|^{x-1}}\frac{\Gamma\left(\frac{1}{|2\kappa|}-\frac{x-1}{2}\right)}{\Gamma\left(\frac{1}{|2\kappa|}+\frac{x-1}{2}\right)}\Gamma(x).
\end{equation}

As for any $\kappa$-deformed function, in the limit $\kappa\to0$ all the $\kappa$-deformed functions tend to their ordinary undeformed expressions, \emph{i.e.}  $\Gamma_{\{0\}}(x)=\Gamma(x)$.\\

This paper is organized as follows: In Section 2 we derive what we call the $\kappa$-Generalized Gamma ($\kappa$GG) distribution and obtain some characteristic statistical properties and inequality measures. In Section 3 the $\kappa$GG distribution tested out by simulating a kinetic exchange model with individual saving propensity. In Section 4 we include a summary and discussion.
\section{The $\kappa$-Generalized-Gamma distribution}
The principle of maximum entropy\cite{JaynesPOME1,JaynesPOME2} provides a constructive criterion for setting up the most unbiased probability distribution that can be assigned to a statistical system under certain fixed conditions that maximize the entropy $S$. As defined by Shannon\cite{ShannonIM1,ShannonIM2}: 
\begin{equation}
 S=-\int_0^\infty f(x)~\ln f(x)~dx.
\end{equation}
Like any probability distribution, $f(x)$ must be normalized to unity:
\begin{equation}
 \int_0^\infty f(x)~dx=1,
\end{equation}
with three linearly independent constraints:
\begin{subequations}
\begin{eqnarray}\nonumber
\begin{aligned}
 \int_0^\infty &\ln x ~f(x)~dx= \ln \beta -\frac{1}{\nu}\left[-\psi\left(\frac{\alpha}{\nu}\right)+\frac{1}{2}\psi\left(\frac{1}{2\kappa}-\frac{\alpha-\nu}{2\nu}\right)\right.\\
 &\left.+\frac{1}{2}\psi\left(\frac{1}{2\kappa}+\frac{\alpha-\nu}{2\nu}\right)+\ln(2\kappa)+\frac{\kappa\nu}{\nu+\kappa(\alpha-\nu)}\right],
\end{aligned} 
\end{eqnarray}
\begin{eqnarray}\nonumber
\begin{aligned}
 \int_0^\infty &\ln\left[1+\kappa^2\left(\frac{x}{\beta}\right)^{2\nu}\right] f(x)dx=\frac{1-\kappa^2\left(\frac{\alpha}{\nu}-1\right)^2}{\left(2\kappa\right)^{\frac{\alpha}{\nu}-1}\Gamma_{\{\kappa\}}\left(\frac{\alpha}{\nu}\right)}\sum_{j=0}^\infty(-1)^{j}{\frac{\alpha}{\nu}-1\choose j}\\
 &\times\left\{\frac{2\kappa}{1-\kappa\left(\frac{\alpha}{\nu}-1-2j\right)}-\psi\left(\frac{2j+5}{4}-\frac{\alpha}{4\nu}+\frac{1}{4\kappa}\right)\right.\nonumber\\
 &\left.+\psi\left(\frac{2j+3}{4}-\frac{\alpha}{4\nu}+\frac{1}{4\kappa}\right)\right\},
\end{aligned}
\end{eqnarray}
and:
\begin{eqnarray}\nonumber
\begin{aligned}
 \int_0^\infty& \ln\left[\sqrt{1+\kappa^2\left(\frac{x}{\beta}\right)^{2\nu}}-\kappa\left(\frac{x}{\beta}\right)^{2\nu}\right] f(x)dx\\
 &=\frac{\kappa}{\left(2\kappa\right)^{\frac{\alpha}{\nu}-1}\Gamma_{\{\kappa\}}\left(\frac{\alpha}{\nu}\right)}\sum_{j=0}^{\infty}(-1)^{j+1}{\frac{\alpha}{\nu}-1\choose j}\frac{1-\kappa^2\left(\frac{\alpha}{\nu}-1\right)^2}{\left[1-\kappa\left(\frac{\alpha}{\nu}-1-2j\right)\right]^2}.
\end{aligned} 
\end{eqnarray}
\end{subequations}
The solution to this maximization problem is given by the following probability distribution:
\begin{eqnarray}\nonumber
\begin{aligned}
\label{kgg}
 f&(x|\alpha,\nu,\beta,\kappa)\\
 &\equiv\left[1-\kappa^2\left(\frac{\alpha}{\nu}-1\right)^2\right]\frac{\nu/\beta}{\Gamma_{\{\kappa\}}(\frac{\alpha}{\nu})}\left(\frac{x}{\beta}\right)^{\alpha-1}\frac{\exp_{\{\kappa\}}\left[-\left(\frac{x}{\beta}\right)^{\nu}\right]}{\sqrt{1+\kappa^2\left(\frac{x}{\beta}\right)^{2\nu}}},
 \end{aligned}
\end{eqnarray}
defined for every $~x\geq0$, where $\alpha,\nu,\beta>0$ are the distribution parameters and $\kappa\in[0,1]$ is a deformation parameter. The probability density function defined by (\ref{kgg}) is what we call the $\kappa$-Generalized-Gamma ($\kappa$GG) distribution since in the limit $\kappa\to0$ it approaches a three parameter Generalized Gamma distribution\cite{Amoroso1925,Stacy1962}. In the same direction as for the Generalized Gamma model containing a family of distributions obtained as special cases such as the Exponential, Gamma or Weibull distributions, the $\kappa$GG distribution can also be parametrized to obtain $\kappa$-deformed versions of these models. In fact, by setting $\nu=\alpha=1$ and $\nu=1$, one obtains a $\kappa$-Exponential distribution and a $\kappa$-Gamma distribution respectively. Moreover, if we set $\nu=\alpha$ in (\ref{kgg}) the $\kappa$-Generalized distribution\cite{Clementi} follows directly:
\begin{equation}
 \label{kgeneralized}
 f(x|\alpha,\beta,\kappa)=\frac{\alpha}{\beta}\left(\frac{x}{\beta}\right)^{\alpha-1}\frac{\exp_{\{\kappa\}}\left[-\left(\frac{x}{\beta}\right)^{\alpha}\right]}{\sqrt{1+\kappa^2\left(\frac{x}{\beta}\right)^{2\alpha}}},
\end{equation}
which is precisely a $\kappa$-deformation of the Weibull distribution.
%%%%%%%%%%%%%%%%%%%%%%%%%%%%%%%%%
\subsection{Tails of the distribution}
Taking the limit $x\to0^+$, the $\kappa$GG distribution (\ref{kgg}) behaves similarly to the ordinary Generalized Gamma distribution:
\begin{eqnarray}\nonumber
% \label{kgg}
 \lim_{x\to0^+}f(x|\alpha,\nu,\beta,\kappa)\sim x^{\alpha-1},
\end{eqnarray}
while for large values of $x$ it approaches a power-law distribution:
\begin{equation*}
 \lim_{x\to\infty}f(x|\alpha,\nu,\beta,\kappa)=\frac{a(x_0)^a}{x^{a+1}},
\end{equation*}
with a characteristic scale:
\begin{equation}
\label{paretoscale}
 x_0=\beta\left[\frac{1+\kappa\left(\frac{\alpha}{\nu}-1\right)}{\Gamma_{\{\kappa\}}\left(\frac{\alpha}{\nu}\right)}~(2\kappa)^{-\frac{1}{\kappa}}\right]^{\frac{\kappa}{\nu\left[1-\kappa\left(\frac{\alpha}{\nu}-1\right)\right]}},
\end{equation}
with $\Gamma_{\{\kappa\}}(\cdot)$ given by (\ref{kgammaf}), and the Pareto exponent:
\begin{equation}
\label{paretoshape}
 a=\frac{\nu}{\kappa}\left[1-\kappa\left(\frac{\alpha}{\nu}-1\right)\right],
\end{equation}
in agreement with the weak Pareto law\cite{Mandelbrot60}.
%%%%%%%%%%%%%%%%%%%%%%%%%%%%%%%%%
\subsection{Elementary properties}
Let:
\begin{equation}
 \label{fggCFD0}
 F(x|\alpha,\nu,\beta,\kappa)=\int_0^x f(t|\alpha,\nu,\beta,\kappa)~dt,
\end{equation}
be the cumulative distribution function of the $\kappa$GG density, then:
\begin{equation}
\label{fggCFD}
 F(x|\alpha,\nu,\beta,\kappa)=1-I_X\left(\frac{1}{2\kappa}-\frac{\alpha-\nu}{2\nu};\frac{\alpha}{\nu}\right),
\end{equation}
with $I_X(\cdot,\cdot)$ is the regularized incomplete beta function and $X=\left(\exp_{\{\kappa\}}[-(x/\beta)^\nu]\right)^{2\kappa}$. Given the form of (\ref{fggCFD}) the median of the $\kappa$-GG distribution does not have a closed form equation since is not possible to find a solution to the equation $F(x_{med})=1/2$, therefore the median of the $\kappa$GG distribution can not be determined analytically, yet the mode can be directly determined by maximizing the distribution:
\begin{eqnarray}\nonumber
\begin{aligned}
 x_{mode}=&\beta\left[\frac{\nu^2+2\kappa^2(\alpha-1)(1+\nu-\alpha)}{2\kappa^2[\nu^2-\kappa^2(1+\nu-\alpha)^2]}\right]^{\frac{1}{2\nu}}\\
 &\times\left\{\sqrt{1+\frac{4\kappa^2[\nu^2-\kappa^2(1+\nu-\alpha)^2](\alpha-1)^2}{[\nu^2+2\kappa^2(\alpha-1)(1+\nu-\alpha)]^2}}-1\right\}^{\frac{1}{2\nu}}.
 \end{aligned}
\end{eqnarray}
% \subsection{Special cases}
\subsection{Moments and statistical measures}\label{moments}
The $r$th-order moment about the origin of the $\kappa$GG distribution:
\begin{eqnarray}\nonumber
% \begin{aligned}
 \mu'_r=\int_0^\infty x^rf(x|\alpha,\beta,\nu,\kappa)dx,
%  \end{aligned}
\end{eqnarray}
writes as:
\begin{eqnarray}
% \begin{aligned}
\label{kggmoments}
 \mu'_r=\beta^r(2\kappa)^{-\frac{r}{\nu}}\frac{1+\kappa\left(\frac{\alpha}{\nu}-1\right)}{1+\kappa\left(\frac{\alpha+r}{\nu}-1\right)}\frac{\Gamma\left(\frac{1}{2\kappa}-\frac{\alpha-\nu+r}{2\nu}\right)}{\Gamma\left(\frac{1}{2\kappa}+\frac{\alpha-\nu+r}{2\nu}\right)}\frac{\Gamma\left(\frac{1}{2\kappa}+\frac{\alpha-\nu}{2\nu}\right)}{\Gamma\left(\frac{1}{2\kappa}-\frac{\alpha-\nu}{2\nu}\right)}\frac{\Gamma\left(\frac{\alpha+r}{\nu}\right)}{\Gamma\left(\frac{\alpha}{\nu}\right)},
%  \end{aligned}
\end{eqnarray}
and exists for $-\alpha<r<\nu/\kappa+(\nu-\alpha)$.
Clearly, by setting $r=1$ in (\ref{kggmoments}) the mean of the distribution $\mu'_1\equiv m$ is directly obtained:
\begin{eqnarray}
% \begin{aligned}
% \label{kggmoments}
 m=\beta(2\kappa)^{-\frac{1}{\nu}}\frac{1+\kappa\left(\frac{\alpha}{\nu}-1\right)}{1+\kappa\left(\frac{\alpha+1}{\nu}-1\right)}\frac{\Gamma\left(\frac{1}{2\kappa}-\frac{\alpha-\nu+1}{2\nu}\right)}{\Gamma\left(\frac{1}{2\kappa}+\frac{\alpha-\nu+1}{2\nu}\right)}\frac{\Gamma\left(\frac{1}{2\kappa}+\frac{\alpha-\nu}{2\nu}\right)}{\Gamma\left(\frac{1}{2\kappa}-\frac{\alpha-\nu}{2\nu}\right)}\frac{\Gamma\left(\frac{\alpha+1}{\nu}\right)}{\Gamma\left(\frac{\alpha}{\nu}\right)}.
%  \end{aligned}
\end{eqnarray}

\begin{figure}[htb!]
 \centering
 \begin{subfigure}[htb!]{0.48\textwidth}
 \includegraphics[width=\textwidth]{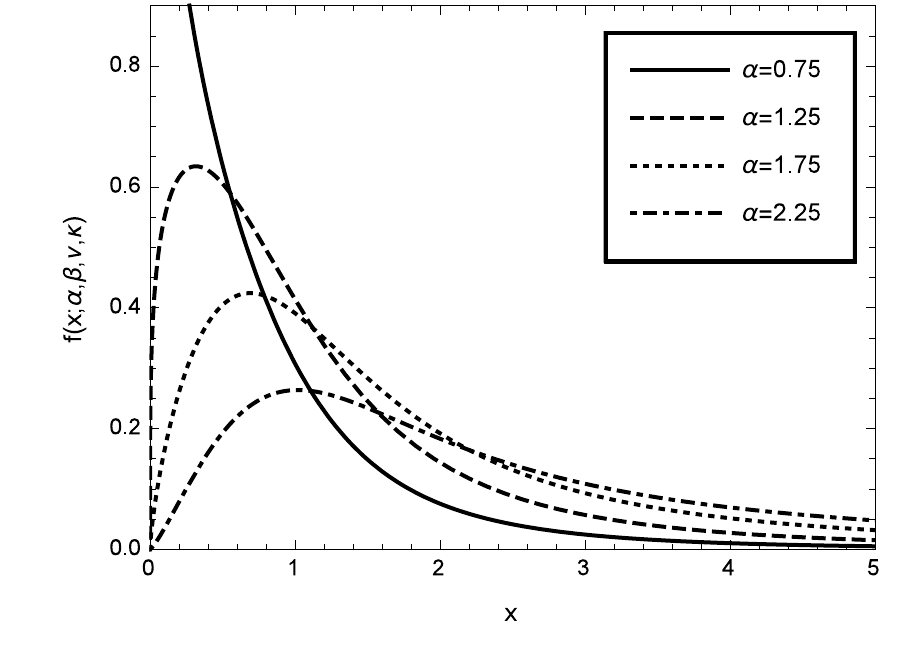}
%  \caption{}\label{fig:Figure1a}
 \end{subfigure}
%  \addtocounter{figure}{-1}
 \hfill
 \begin{subfigure}[htb!]{0.48\textwidth}
 \includegraphics[width=\textwidth]{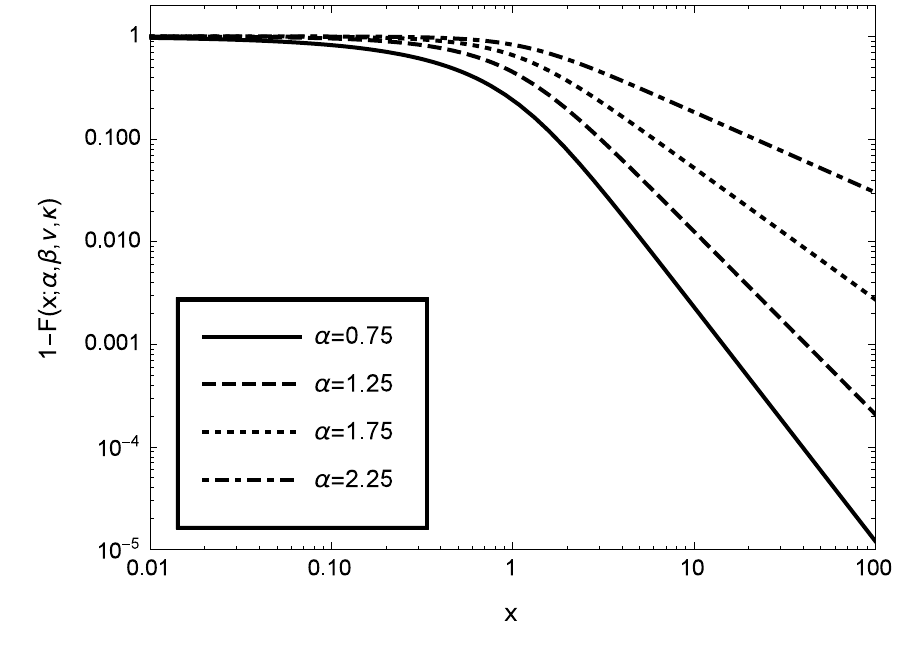}
%  \caption{(b)}  \label{fig:Figure1b}
 \end{subfigure}
 \caption{$\kappa$GG distribution functions (left) and complementary cumulative distributions (right) for $\beta=1.20,~\nu=1.3,~\kappa=0.75$ fixed and different values of $\alpha$.}
 \label{fig:Figure1}
\end{figure}

\begin{figure}[htb!]
 \centering
 \begin{subfigure}[b]{0.48\textwidth}
 \includegraphics[width=\textwidth]{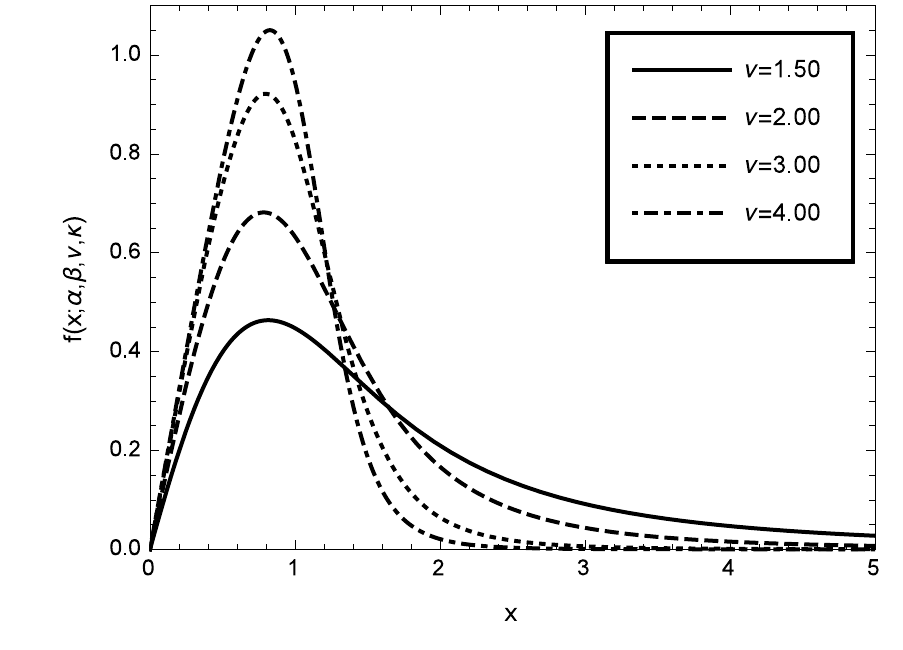}
%  \caption{(a)}
 \end{subfigure}
%  \addtocounter{figure}{-1}
 ~
 \begin{subfigure}[b]{0.48\textwidth}
 \includegraphics[width=\textwidth]{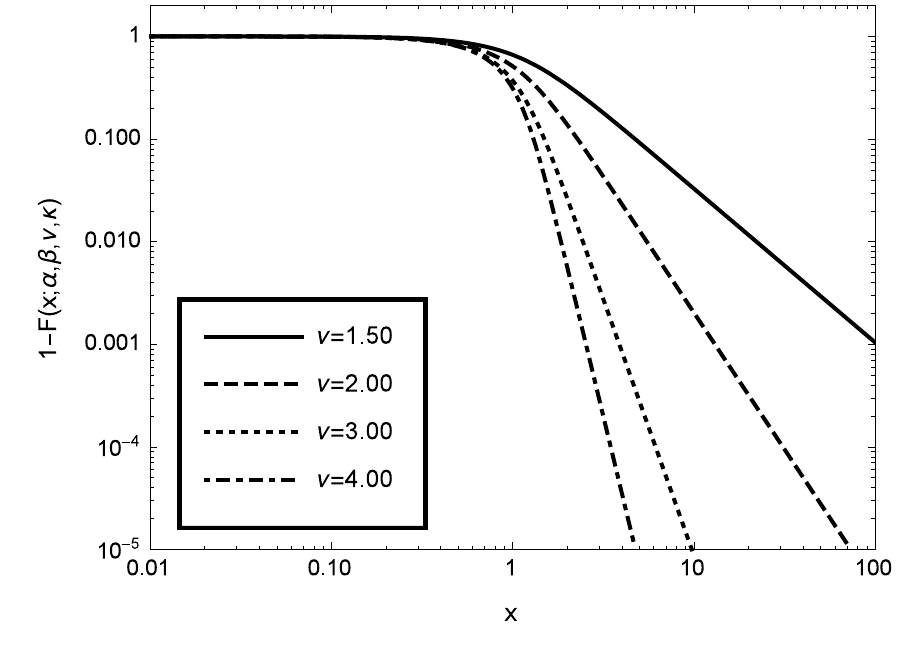}
%  \caption{(b)}
 \end{subfigure}
 \caption{\label{Figure2}$\kappa$GG distribution functions (left) and complementary cumulative distributions (right) for $\alpha=2.00,~\beta=1.20,~\kappa=0.75$ fixed and different values of $\nu$.}
\end{figure}

\begin{figure}[htb!]
 \centering
 \begin{subfigure}[b]{0.48\textwidth}
 \includegraphics[width=\textwidth]{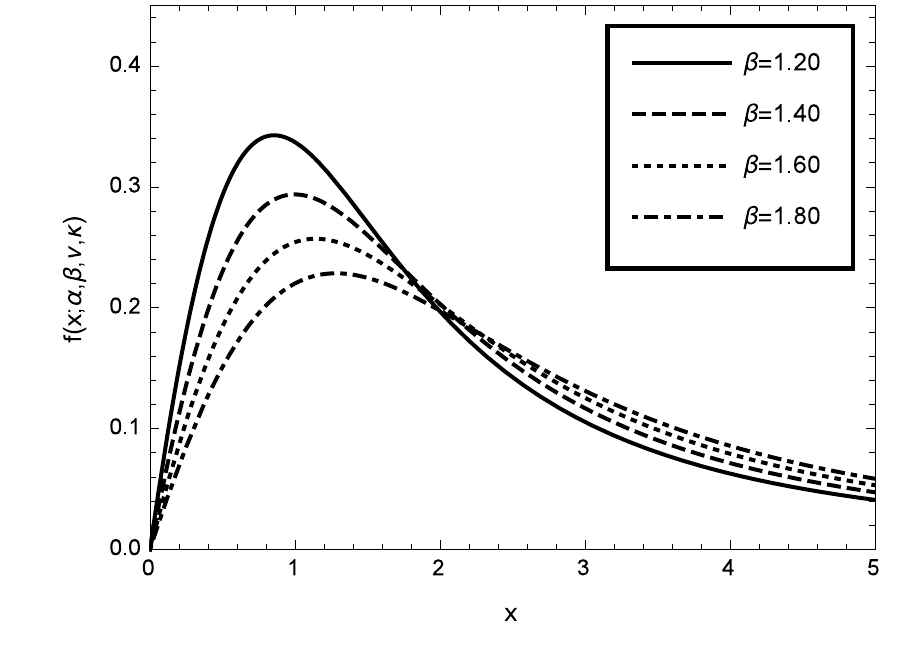}
%  \caption{(a)}
 \end{subfigure}
%  \addtocounter{figure}{-1}
 ~
 \begin{subfigure}[b]{0.48\textwidth}
 \includegraphics[width=\textwidth]{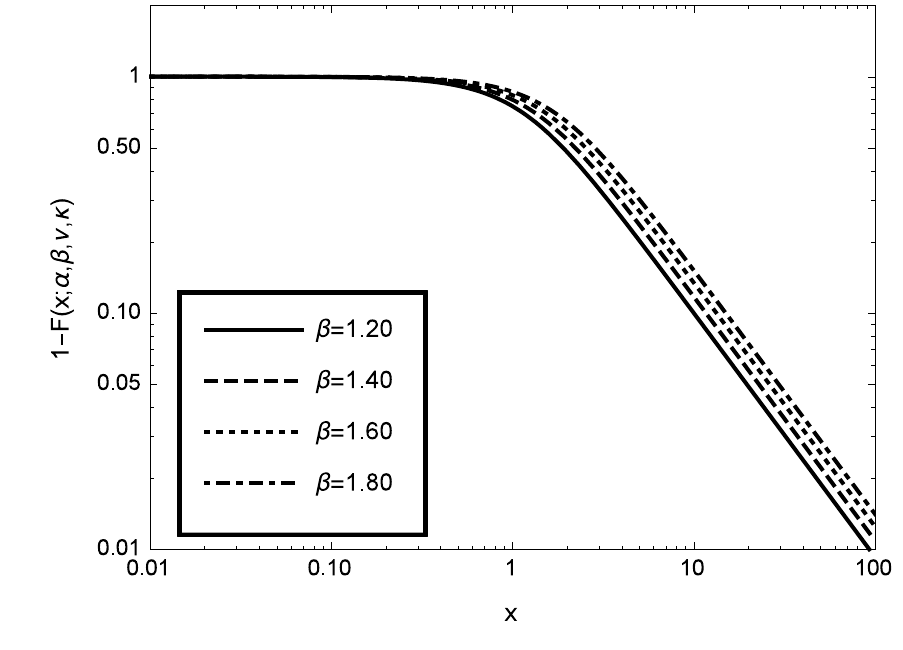}
%  \caption{(b)}
 \end{subfigure}
 \caption{\label{Figure3}$\kappa$GG distribution functions (left) and complementary cumulative distributions (right) for $\alpha=2.00,~\nu=1.30,~\kappa=0.75$ fixed and different values of $\beta$.}
\end{figure}

Analogously, the variance relates directly with the moments as:
\begin{eqnarray}
\begin{aligned}
 \sigma^2=&\mu'_2-m^2\nonumber\\
 =&\beta^2\frac{1+\kappa\left(\frac{\alpha}{\nu}-1\right)}{(2\kappa)^{\frac{2}{\nu}}\Gamma\left(\frac{\alpha}{\nu}\right)}\frac{\Gamma\left(\frac{1}{2\kappa}+\frac{\alpha-\nu}{2\nu}\right)}{\Gamma\left(\frac{1}{2\kappa}-\frac{\alpha-\nu}{2\nu}\right)}\left\{\frac{\Gamma\left(\frac{\alpha+2}{\nu}\right)}{1+\kappa(\frac{\alpha+2}{\nu}-1)}\frac{\Gamma\left(\frac{1}{2\kappa}-\frac{\alpha-\nu+2}{2\nu}\right)}{\Gamma\left(\frac{1}{2\kappa}+\frac{\alpha-\nu+2}{2\nu}\right)}\right.\nonumber\\
 &\left.-\frac{1+\kappa\left(\frac{\alpha}{\nu}-1\right)}{\Gamma\left(\frac{\alpha}{\nu}\right)}\frac{\Gamma\left(\frac{1}{2\kappa}+\frac{\alpha-\nu}{2\nu}\right)}{\Gamma\left(\frac{1}{2\kappa}-\frac{\alpha-\nu}{2\nu}\right)}\left[\frac{\Gamma\left(\frac{\alpha+1}{\nu}\right)}{1+\kappa(\frac{\alpha+1}{\nu}-1)}\frac{\Gamma\left(\frac{1}{2\kappa}-\frac{\alpha-\nu+1}{2\nu}\right)}{\Gamma\left(\frac{1}{2\kappa}+\frac{\alpha-\nu+1}{2\nu}\right)}\right]^2\right\},
\end{aligned}
\end{eqnarray}
consequently, the coefficient of variation or relative variability corresponds to the ratio of the standard deviation to the mean:
\begin{equation}
CV=\frac{\sigma}{m}=\sqrt{\frac{\frac{\Gamma\left(\frac{\alpha+2}{\nu}\right)}{1+\kappa(\frac{\alpha+2}{\nu}-1)}\frac{\Gamma\left(\frac{1}{2\kappa}-\frac{\alpha-\nu+2}{2\nu}\right)}{\Gamma\left(\frac{1}{2\kappa}+\frac{\alpha-\nu+2}{2\nu}\right)}}{\frac{1+\kappa\left(\frac{\alpha}{\nu}-1\right)}{\Gamma\left(\frac{\alpha}{\nu}\right)}\left[\frac{\Gamma\left(\frac{\alpha+1}{\nu}\right)}{1+\kappa(\frac{\alpha+1}{\nu}-1)}\frac{\Gamma\left(\frac{1}{2\kappa}-\frac{\alpha-\nu+1}{2\nu}\right)}{\Gamma\left(\frac{1}{2\kappa}+\frac{\alpha-\nu+1}{2\nu}\right)}\right]^2}-1}.
\end{equation}
% \newpage
%%%%%%%%%%%%%%%%%%%%%%%%%%%%%%%%%%%%%%%%%%%%%%%%%%%%%%%%%%%%%%%%%
The $\kappa$GG distribution is completely characterized by four real parameters: a deformation parameter $\kappa$, a scale parameter $\beta$ and two shape parameters $\alpha$ and $\nu$. Figure \ref{fig:Figure1} shows that as $\alpha$ becomes smaller the distribution turns more concentrated around the $x=0^+$ border and is mostly determined by the exponential part of the distribution, whereas for larger values it becomes more spread out. As $\nu$ becomes larger in Figure \ref{Figure2} the distribution stretches vertically and starts to concentrate around the mode, whereas for smaller values it also becomes more spread out. The scale parameter $\beta$ in Figure \ref{Figure3} measures the statistical dispersion of the distribution, that is, for small (large) values of $\beta$ the distribution gets more stretched out (squeezed). Finally, the deformation parameter $\kappa$ does not produce any significant effect on the shape of the distribution as it affects the tail, the larger (smaller) the $\kappa$ the heavier (lighter) the tail of the distribution. Note that, indirectly, the tail of the distribution is also affected by the shape parameters, becoming heavier by increasing $\alpha$ and decreasing $\nu$ or viceversa.
% \newpage
% % %%%%%%%%%%%%%%%%%%%%%%%%%%%%%%%%%%%%%%%%%%%%%%%%%%%%%%%%%%%%%
% Figure2%%%%%%%%%%%%%%%%%%%%%%%%%%%%%%%%%%%%%%%%%%%%%%%%%%%%%%%%
% Figure3%%%%%%%%%%%%%%%%%%%%%%%%%%%%%%%%%%%%%%%%%%%%%%%%%%%%%%%%%%%
% Figure4%%%%%%%%%%%%%%%%%%%%%%%%%%%%%%%%%%%%%%%%%%%%%%%%%%%%%%%%
\begin{figure}[htb!]
 \centering
 \begin{subfigure}[b]{0.48\textwidth}
 \includegraphics[width=\textwidth]{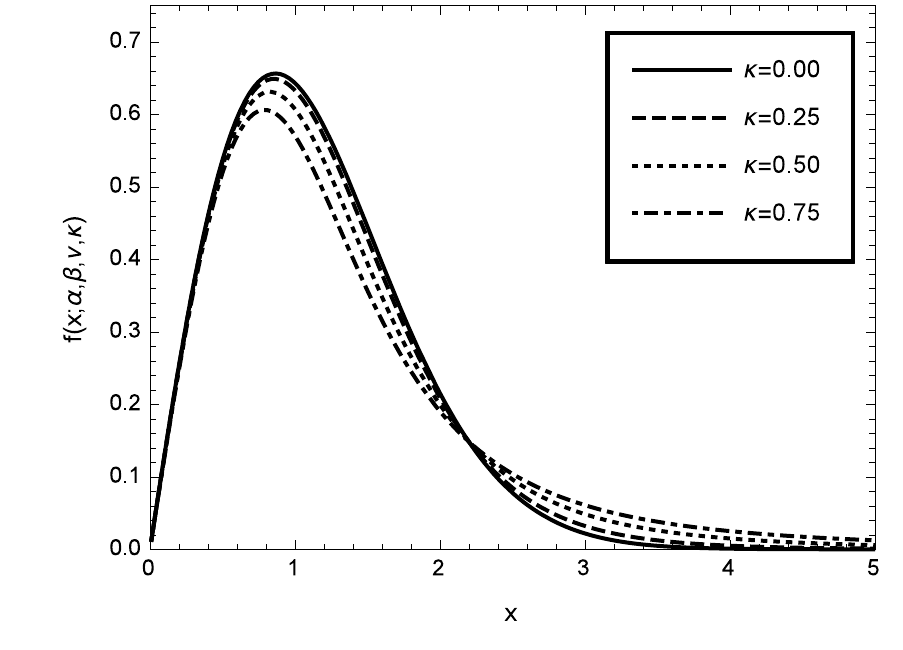}
%  \caption{(a)}
 \end{subfigure}
%  \addtocounter{figure}{-1}
 ~
 \begin{subfigure}[b]{0.48\textwidth}
 \includegraphics[width=\textwidth]{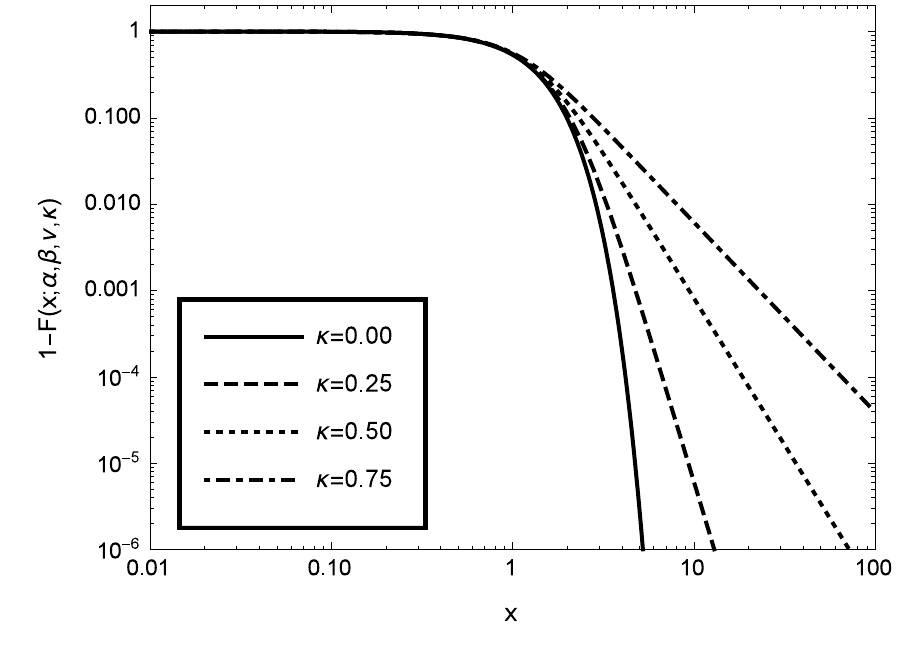}
%  \caption{(b)}
 \end{subfigure}
 \caption{\label{Figure4}$\kappa$GG distribution functions (left) and complementary cumulative distributions (right) for $\alpha=2.00,~\beta=1.20,~\nu=1.80$ fixed and different values of  $\kappa$.}
\end{figure}
%%%%%%%%%%%%%%%%%%%%%%%%%%%%%%%%%%%%%%%%%%%%%%%%%%%%%%%%%%%%%%%%%%%%%%%
\subsection{Lorenz Curve and inequality measures}
The Lorenz curve\cite{Lorenz1905,Gastwirth1971} for the $\kappa$GG distribution in terms of the probability density function (\ref{kgg}) can be expressed analytically as: 
\begin{eqnarray}
\label{kgglorenz}
\begin{aligned}
 L(F(x))=&\frac{1}{m}\int_0^xt~ f(t|\alpha,\beta,\nu,\kappa)~dt\nonumber\\
 =&1-\frac{1+\kappa\left(\frac{1+\alpha-\nu}{\nu}\right)}{2\Gamma\left(\frac{1+\alpha-\nu}{\nu}\right)}\frac{\Gamma\left(\frac{1}{2\kappa}+\frac{1+\alpha-\nu}{2\nu}\right)}{\Gamma\left(\frac{1}{2\kappa}-\frac{1+\alpha-\nu}{2\nu}\right)}\nonumber\\
 &\times \left\{B_X\left(\frac{1}{2\kappa}-\frac{1+\alpha-\nu}{2\nu},\frac{1+\alpha-\nu}{\nu}\right)\right.\nonumber\\
 &+B_X\left(\frac{1}{2\kappa}-\frac{1+\alpha-\nu}{2\nu}+1,\frac{1+\alpha-\nu}{\nu}\right)\nonumber\\
 &+\left.\frac{2\nu}{1+\alpha-\nu} (2\kappa)^{\frac{1+\alpha-\nu}{\nu}}(1-u)\left[\ln_{\kappa}\left(\frac{1}{1-u}\right)\right]^{\frac{1+\alpha-\nu}{\nu}}\right\},
\end{aligned}
\end{eqnarray}
where $B_X(\cdot,\cdot)$ is the incomplete beta function with $X=(1-u)^{2\kappa}$. The Lorenz curve (\ref{kgglorenz}) exists if the condition $\nu/\kappa-(\alpha-\nu)>1$ is satisfied.

\begin{dfn}
 \label{definition1}
 Let $X,Y$ be two non-negative random variables with positive finite expectations, and their associated Lorenz curves $L_X(u)$ and $L_Y(u)$ respectively. The Lorenz partial order, denoted $\leq_L$ defines as:
 \begin{equation}
  X\leq_LY\Longleftrightarrow L_X(u)\geq L_Y(u),~\forall u\in[0,1].
 \end{equation}
\end{dfn}

Definition \ref{definition1}\cite{Arnold1984} simply states that the distribution of $X$ exhibits less inequality in the Lorenz sense than the distribution of $Y$ if the Lorenz curve of $X$ lies above the Lorenz curve of $Y$. Let us now find the Lorenz ordering conditions for any set of $\kappa$GG distributions\cite{Arnold1986,Hardy1929}.
\begin{thm}
\label{Theorem}
 Let $X$ and $Y$ be two $\kappa$GG distributed random variables. Then:
 \begin{equation}
\label{theoremeq}  
X\leq_{L}Y\Longleftarrow:\alpha_x\leq\alpha_y,~and~\frac{\nu_x}{\kappa_x}-(\alpha_x-\nu_x)\leq\frac{\nu_y}{\kappa_y}-(\alpha_y-\nu_y).
 \end{equation}
\end{thm}
 A detailed proof of this theorem can be found in the Appendix.
 
% %% Figure5%%%%%%%%%%%%%%%%%%%%%%%%%%%%%%%%%%%%%%%%%%%%%%%%%%%%%%
\begin{figure}[htb!]
 \centering
 \begin{subfigure}[b]{0.47\textwidth}
 \includegraphics[width=\textwidth]{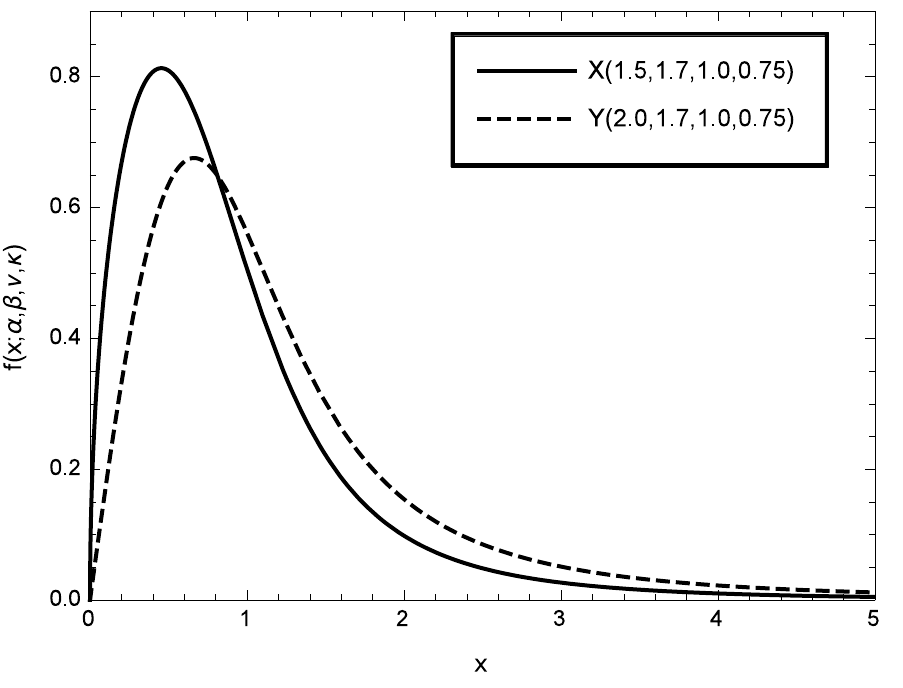}
%  \caption{(a)}
 \end{subfigure}
%  \addtocounter{figure}{-1}
 ~
 \begin{subfigure}[b]{0.47\textwidth}
 \includegraphics[width=\textwidth]{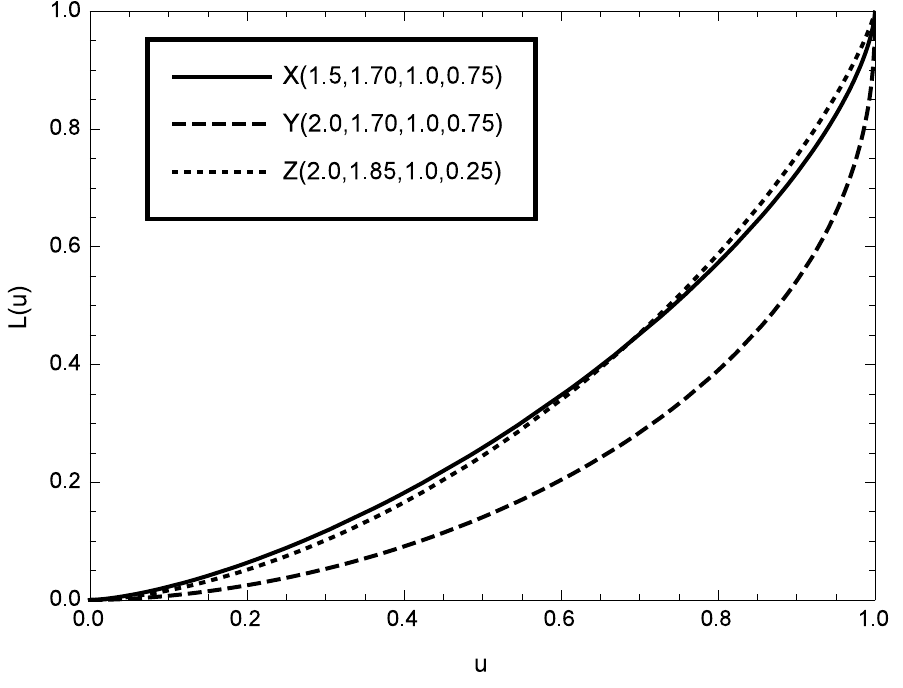}
%  \caption{(b)}
 \end{subfigure}
 \caption{\label{Figure5} If the parameters of any set of $\kappa$GG functions (left) are selected accordingly to theorem \ref{Theorem} then their Lorenz curves (right) are partially ordered in the Lorenz sense, otherwise they intersect as the dotted line shows.}
\end{figure}
%%%%%%%%%%%%%%%%%%%%%%%%%%%%%%%%%%%%%%%%%%%%%%%%%%%%%%%%%%%%%%%%%%%
Let us observe the left hand side of Figure \ref{Figure5}. For any given set of $\kappa$GG distributions, the solid line represents in principle a less unequal distribution since it is more concentrated around the mode and it has a lighter tail than the dashed line. In the Lorenz sense, the right hand side shows that when the parameters are chosen according to theorem \ref{Theorem} the Lorenz curve of the less unequal distribution lies above that of the more unequal one. Therefore $L_X\geq L_Y$ and $L_Z\geq L_Y$. Nevertheless, if the parameters are not chosen according to theorem \ref{Theorem} then the distributions are not comparable. In other words, $X$ and $Z$ are less unequal than $Y$ but nothing can be said about the Lorenz ordering between $X$ and $Z$ since their parameters do not satisfy the prescribed condition.\\

In terms of inequality, there exist many ways to determine the dispersion of incomes. Amongst the most known metrics, the Gini coefficient\cite{Gini1914} is also based on the Lorenz curve and can be determined in terms of a probability density function $f$ as: 
\begin{equation*}
 G=1-\frac{2}{m}\int_0^\infty f(x)\int_0^xt~f(t)~dt~dx,
\end{equation*}
where $m$ is the mean of the distribution, then:
\begin{eqnarray}
 \label{equation25}
 G=1-&\frac{1}{\nu}\left[\nu+\kappa\left(\alpha-\nu\right)\right]\frac{1+\kappa\left(\frac{\alpha-\nu+1}{\nu}\right)}{1+\kappa\left(\frac{2\alpha-2\nu+1}{2\nu}\right)}\frac{\Gamma\left(\frac{1}{2\kappa}+\frac{\alpha-\nu}{2\nu}\right)}{\Gamma\left(\frac{1}{2\kappa}-\frac{\alpha-\nu}{2\nu}\right)}\nonumber\\
 &\times\frac{\Gamma\left(\frac{1}{\kappa}-\frac{2\alpha-2\nu+1}{2\nu}\right)}{\Gamma\left(\frac{1}{\kappa}+\frac{2\alpha-2\nu+1}{2\nu}\right)}\frac{\Gamma\left(\frac{1}{2\kappa}+\frac{\alpha-\nu+1}{2\nu}\right)}{\Gamma\left(\frac{1}{2\kappa}-\frac{\alpha-\nu+1}{2\nu}\right)}\frac{\Gamma\left(\frac{2\alpha-\nu+1}{\nu}\right)}{\Gamma\left(\frac{\alpha+1}{\nu}\right)\Gamma\left(\frac{\alpha}{\nu}\right)}.
\end{eqnarray}

Another family of inequality measures is the generalized entropy \cite{Cowell1980a,Cowell1980b,Shorrocks1980,CowellKuga1981a,CowellKuga1981b}:
\begin{eqnarray}\nonumber
% \begin{aligned}
 \label{equation26}
 GE(\theta)=\frac{1}{\theta^2-\theta}&\left\{\left(\frac{\beta}{m}\right)^\theta\left[(2\kappa)^{-\frac{\theta}{\nu}}\frac{\nu+\kappa(\alpha-\nu)}{\nu+\kappa(\theta+\alpha-\nu)}\right.\right.\\
 &\times\left.\left.\frac{\Gamma\left(\frac{1}{2\kappa}-\frac{\theta+\alpha-\nu}{2\nu}\right)}{\Gamma\left(\frac{1}{2\kappa}+\frac{\theta+\alpha-\nu}{2\nu}\right)}\frac{\Gamma\left(\frac{1}{2\kappa}-\frac{\alpha-\nu}{2\nu}\right)}{\Gamma\left(\frac{1}{2\kappa}+\frac{\alpha-\nu}{2\nu}\right)}\frac{\Gamma\left(\frac{\theta+\alpha}{2\nu}\right)}{\Gamma\left(\frac{\alpha}{2\nu}\right)}\right]-1\right\}.
% \end{aligned}
\end{eqnarray}
Note that the generalized entropy (\ref{equation26}) can assume different forms depending on the value of $\theta$. Firstly, the mean logarithmic deviation is obtained by taking the limit $\theta\to0$ in the generalized entropy measure:
\begin{eqnarray}\nonumber
% \begin{aligned}
 \label{equation27}
 MLD= \frac{1}{\nu}&\left[-\psi\left(\frac{\alpha}{\nu}\right)+\frac{1}{2}\psi\left(\frac{1}{2\kappa}-\frac{\alpha-\nu}{2\nu}\right)+\frac{1}{2}\psi\left(\frac{1}{2\kappa}+\frac{\alpha-\nu}{2\nu}\right)\right.\\
 &\left.\ln(2\kappa)-\nu\ln\left(\frac{\beta}{m}\right)+\frac{\kappa\nu}{\nu+\kappa(\alpha-\nu)}\right],
% \end{aligned}
\end{eqnarray}
where $\psi(z)=\Gamma'(z)/\Gamma(z)$ is the Digamma function. Secondly, the Theil\cite{Theil1967} index can be obtained by taking the limit $\theta\to1$ in the generalized entropy measure:
 \begin{eqnarray}\nonumber
%  \begin{aligned}
  \label{equation28}
  T= \frac{1}{\nu}&\left[\psi\left(\frac{\alpha+1}{\nu}\right)-\frac{1}{2}\psi\left(\frac{1}{2\kappa}-\frac{\alpha+1-\nu}{2\nu}\right)-\frac{1}{2}\psi\left(\frac{1}{2\kappa}+\frac{\alpha+1-\nu}{2\nu}\right)\right.\\
 &\left.-\ln(2\kappa)+\nu\ln\left(\frac{\beta}{m}\right)-\frac{\kappa\nu}{\nu+\kappa(\alpha+1-\nu)}\right].
%   \end{aligned}
 \end{eqnarray}
The previously derived inequality measures are, as mentioned, amongst the most used and most known ones, and many other interesting metrics could also be calculated but the purpose of this paper is not on that direction.\\

As pointed out in previous sections, the $\kappa$GG distribution given by expression (\ref{kgg}) is in fact an extension to the $\kappa$-Generalized model proposed by Clementi et al.\cite{Clementi} since it is a limiting case as the shape parameters $\alpha$ and $\nu$ become equal. The interested reader can easily verify that all previous calculations presented here are consistent to those for the $\kappa$-Generalized model by setting $\nu=\alpha$.
%%%%%%%%%%%%%%%%%%%%%%%%%%%%%%%%%%%%%%%%%%%%%%%%%%%%%%%%%%%%%%%%%%%%%%
\section{Application to a kinetic exchange model}
\begin{figure}[htb!]
 \centering
 \begin{subfigure}[b]{0.47\textwidth}
 \includegraphics[width=\textwidth]{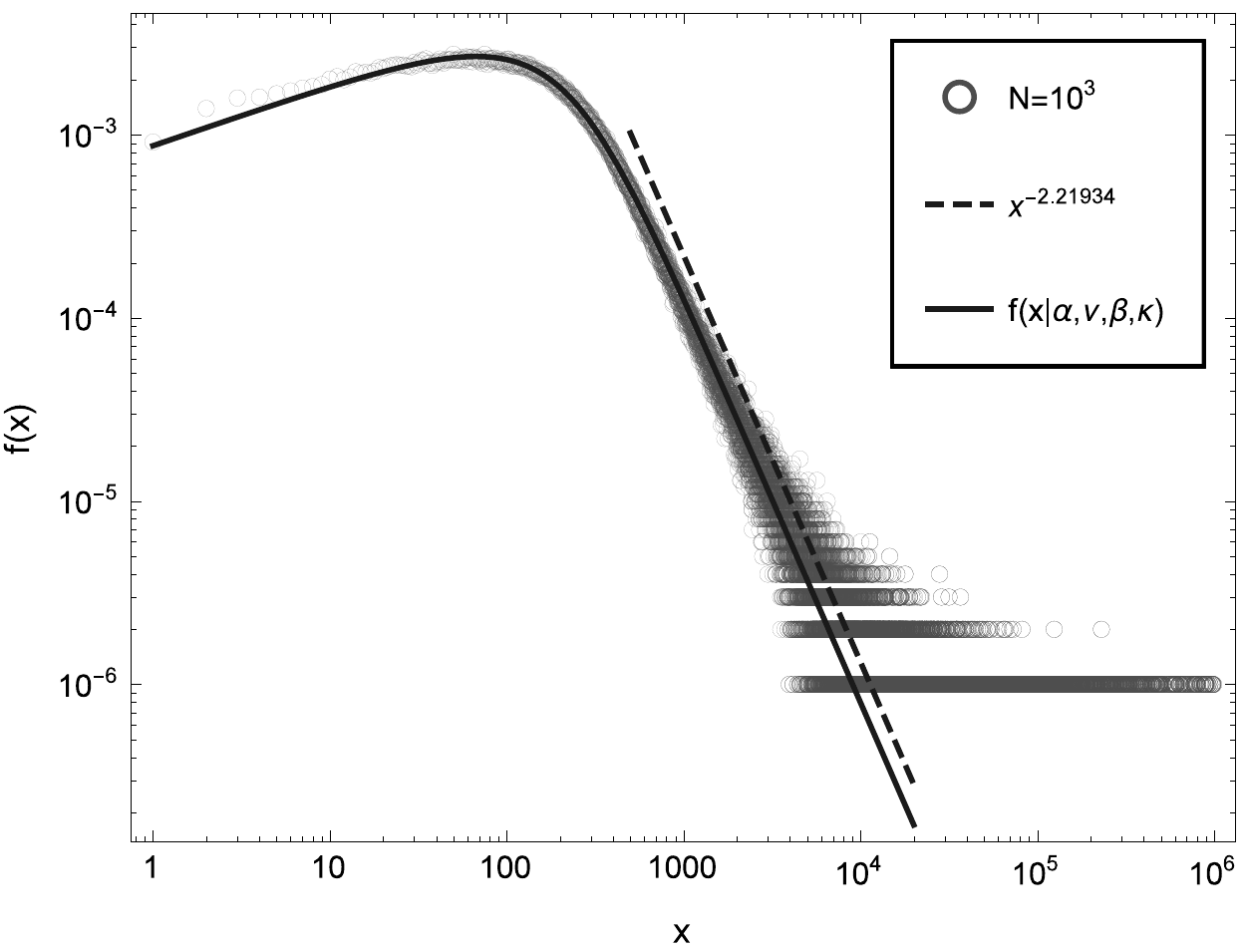}
 \caption{}
 \end{subfigure}
%  \addtocounter{figure}{-1}
 ~
 \begin{subfigure}[b]{0.47\textwidth}
 \includegraphics[width=\textwidth]{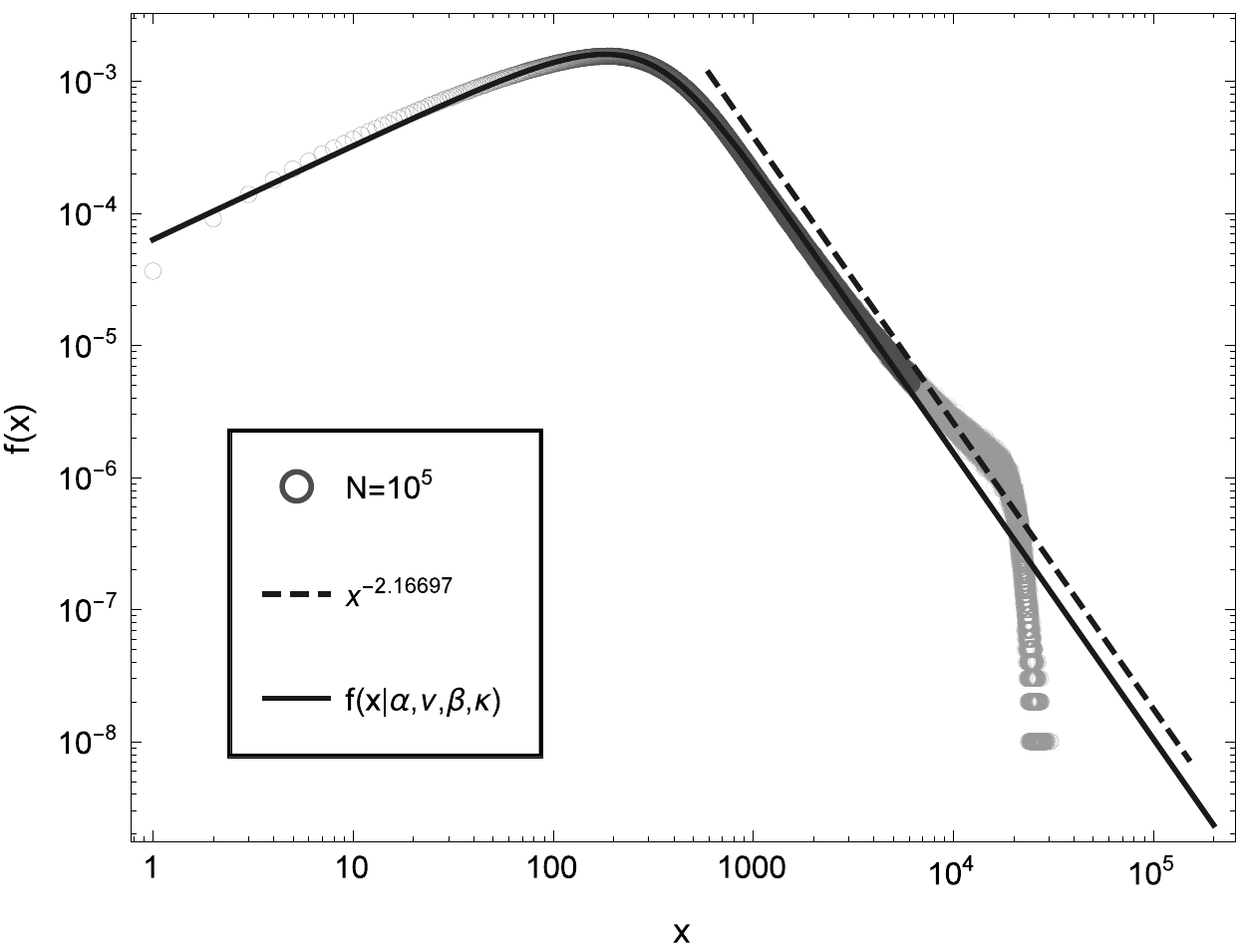}
 \caption{}
 \end{subfigure}
 \caption{\label{Figure6} (a) Steady state distribution of wealth in heterogeneous kinetic exchange models for  $N=10^3$ and (b) $N=10^5$ agents, with distributed saving propensity and average money $M/N=10^{3}$. The solid lines correspond to the fitted $\kappa$GG distributions and the dashed lines are the corresponding power-law tails.}
\end{figure}
In order to test our model we consider the kinetic exchange model with heterogeneous saving propensity investigated by Chatterjee et al. \cite{Chatterjee2003,Chatterjee2004,Bhattacharya2005}. We simulate a closed economic system with total money $M$ and fixed number of agents $N$. Initially the money $M$ is equally distributed among agents, each with an uniformly distributed personal saving parameter $\lambda_i$. At each time, two agents are randomly selected and they exchange money according to equations (\ref{KEM1}) where the trading term given by equation (\ref{heterogeneous}). This procedure is repeated many times until a stable money distribution is observed. In Figure \ref{Figure6} (a)  we show the steady state distribution of wealth $f(x)$ vs $x$ for $N=10^3$ agents and in Figure \ref{Figure6} (b) for $N=10^5$ agents, both with average money $M/N=10^{3}$ after $10^9$ iterations and averaging over $10^3$ realizations. In both case scenarios there is an initial growth of $f(x)$, which quickly reaches a maximum and then a long-range power-law decay for large values of $x$. The results obtained by these simulations have been discussed in detail  by Patriarca et al.\cite{Patriarca2006}. Firstly, in Figure \ref{Figure6} (a), the large dispersion on the tail of the steady state distribution is primarily due to the size of the system, that is, wealthier agents are so few compared to the total $N$ that there is no sufficient data to get finer statistics on the power-law tail. Secondly, in Figure \ref{Figure6} (b), by increasing the number of agents of the system, an abrupt cutoff on the tail of the distribution appears. Given that the power-law tail is primarily produced by the agents with the larger saving propensities, close to the value one, this cutoff shows up due to the discreteness of $\lambda_i$, as pointed out by Patriarca et al.\cite{Patriarca2005}, as $N$ grows the cutoff tends to disappear.\\
%resulting in more and more agents with a higher saving propensities, there will always be an agent with the largest saving propensity $\lambda_{M}$. The larger the $\lambda_M$, the further the cutoff. So the difficulty of getting a finer power-law tail out of the steady state distribution in heterogeneous kinetic exchange models with as one needs to have agents with saving propensities infinitesimally close to one for the dispersion and 
Taking the previous discussion into account, in Figure \ref{Figure6} (a) we considered the whole body of the steady state distribution to perform the fit of the $\kappa$GG distribution, while in the case discussed in Figure \ref{Figure6} (b) we performed the fit of the $\kappa$GG distribution using the darker circles of the histogram. The distribution parameters found in the case represented by Figure \ref{Figure6} (a) are: $\alpha=1.3320$, $\nu=1.1240$, $\beta=221.3150$ and $\kappa=0.7868$ and Pareto exponent $a=1.2193$. The distribution parameters found for the case represented in Figure \ref{Figure6} (b) are: $\alpha=1.7127$, $\nu=1.3710$, $\beta=342.7610$ and $\kappa=0.9088$ and Pareto exponent $a=1.1670$. It is important to notice that the values we obtained for the Pareto exponent in the power-law tail of the distributions are in the range between one and two, corresponding to the observational range reported in the literature\cite{Chatterjee2005,Das2003,Repetowicz2005}. To this end, we pointed out that the proposed $\kappa$GG distribution function presented here can describe the steady state distribution in the discussed kinetic exchange models and observational wealth distributions in a broader range than other known statistical distributions.
%%%%%%%%%%%%%%%%%%%%%%%%%%%%%%%%%%%%%%%%%%%%%%%%%%%%%%%%%%%%%%%%%%%%%%
\section{Summary and discussion}
The main objective of this research is to provide a close mathematical description of the size distribution of wealth determined by agent-based models in isolated heterogeneous systems. Under the assumption that systems that do not have time reversibility obey a nonlinear dynamics in the Kaniadakis sense and from the principle of maximum entropy, we have obtained a four parameter distribution regarded as a $\kappa$-deformation of the Generalized Gamma distribution, that can describe in a single mathematical expression, the entire range of the distribution of wealth. Since Pareto's work, a considerable number of mathematical functions have been considered as possible models for the distribution of wealth, models that can be obtained as particular parametrizations of the $\kappa$GG distribution presented here. This distribution accurately reproduces a broader range of the distribution of wealth for heterogeneous kinetic exchange models, and the corresponding Pareto exponent for the power-law tail is found to be in the range of previous results. We have also obtained expressions for the moments as well as other useful standard statistical tools for the analysis and measurement of wealth inequality.
Finally, we remark that the proposed $\kappa$GG distribution is a natural generalization of other known statistical distributions such as the $\kappa$-Generalized, Generalized Gamma, Gamma, Weibull and the Exponential, allowing the description of wealth distributions in the broadest range.

%%%%%%%%%%%%%%%%%%%%%%%%%%%%%%%%%%%%%%%%%%%%%%%%%%%%%%%%%%%%%%%%%%%%%%%%%%%%%%%%%%%%%%%%%
\appendix
\section*{Appendix}
\setcounter{section}{1}% %
Since the Lorenz curve is invariant under scale changes, the parameter $\beta$ can be chosen as $1$ in (\ref{kgglorenz}) without any loss of generality. Furthermore, we state the following theorem\cite{Arnold1986,Hardy1929}:
\begin{thm}\label{theorem2}
 Let $X,Y$ be two positive random variables with finite mean $m_x,m_y$ respectively. A necessary an sufficient condition that $X\leq_{L}Y$ is that:
 \begin{equation}
  \label{theorem2eq}
  \mathbb{E}\left[\psi\left(\frac{X}{m_{x}}\right)\right]\leq\mathbb{E}\left[\psi\left(\frac{Y}{m_{y}}\right)\right],
 \end{equation}
for every continuous and convex function $\psi:\mathbb{R}_+\to\mathbb{R}$, for which the expectation exists.
\end{thm}
For theorem \ref{Theorem} we need to prove  that $\alpha_x\leq\alpha_y$ and $\nu_x/\kappa_x-(\alpha_x-\nu_x)\leq\nu_y/\kappa_y-(\alpha_y-\nu_y)$ is a necessary and sufficient condition for the Lorenz-ordering of $\kappa$GG distributions. Consider the family of continuous convex functions:
\begin{equation}
 \Psi(x)=\frac{x^{t+1}-1}{t(t+1)},~x>0,~-\infty<t<\infty,~t\neq-1,0.
\end{equation}
As previous results show\cite{Taillie1981,Clementi}, corresponding to $\Psi$ one can obtain a class of inequality measures:
\begin{eqnarray}
 H_t(X)&=\mathbb{E}\left[\psi\left(\frac{X}{m_{x}}\right)\right]\\
 &=\frac{1}{t(t+1)}\left[\frac{\mathbb{E}(X^{t+1})}{m_x^{t+1}}-1\right],~-\infty<t<\infty,~t\neq-1,0,
\end{eqnarray}
that preserve the Lorenz-ordering. From theorem \ref{theorem2}, it follows that:
\begin{equation}
\label{lemma1b}
 H_t(X)\leq H_t(Y),
\end{equation}
and considering expression (\ref{kggmoments}) it follows that:

\begin{eqnarray}
\label{lordering}
 H_t(X)=\frac{1}{t(t+1)}&\left\{\frac{1+\kappa\left(\frac{\alpha}{\nu}-1\right)}{1+\kappa\left(\frac{t+1+\alpha}{\nu}-1\right)}\frac{\Gamma\left(\frac{1}{2\kappa}-\frac{t+1+\alpha-\nu}{2\nu}\right)}{\Gamma\left(\frac{1}{2\kappa}+\frac{t+1+\alpha-\nu}{2\nu}\right)}\frac{\Gamma\left(\frac{1}{2\kappa}+\frac{\alpha-\nu}{2\nu}\right)}{\Gamma\left(\frac{1}{2\kappa}-\frac{\alpha-\nu}{2\nu}\right)}  \right.\nonumber\\
 &\times\frac{\Gamma\left(\frac{t+1+\alpha}{\nu}\right)}{\Gamma\left(\frac{\alpha}{\nu}\right)}\left[\frac{1+\kappa\left(\frac{1+\alpha}{\nu}-1\right)}{1+\kappa\left(\frac{\alpha}{\nu}-1\right)}\frac{\Gamma\left(\frac{1}{2\kappa}+\frac{1+\alpha-\nu}{2\nu}\right)}{\Gamma\left(\frac{1}{2\kappa}-\frac{1+\alpha-\nu}{2\nu}\right)}  \right.\nonumber\\
 &\left.\left.\times\frac{\Gamma\left(\frac{1}{2\kappa}-\frac{\alpha-\nu}{2\nu}\right)}{\Gamma\left(\frac{1}{2\kappa}+\frac{\alpha-\nu}{2\nu}\right)}\frac{\Gamma\left(\frac{\alpha}{\nu}\right)}{\Gamma\left(\frac{1+\alpha}{\nu}\right)}\right]^{t+1}-1\right\}.
\end{eqnarray}
Recalling that $\lim_{z\to0}\Gamma(z)=+\infty$, we see that:
\begin{equation}
 H_t(X)~\exists~\forall ~t~\in~-\alpha_x-1<t<\frac{\nu_x}{\kappa_x}-(\alpha_x-\nu_x)-1,
\end{equation}
and analogously
\begin{equation}
 H_t(Y)~\exists~\forall ~t~\in~-\alpha_y-1<t<\frac{\nu_y}{\kappa_y}-(\alpha_y-\nu_y)-1.
\end{equation}
Finally, in conjunction with (\ref{lemma1b}), it follows pretty straightforwardly that:
\begin{eqnarray}
 &\alpha_x\leq\alpha_y,\\
 &\frac{\nu_x}{\kappa_x}-(\alpha_x-\nu_x)\leq\frac{\nu_y}{\kappa_y}-(\alpha_y-\nu_y),
\end{eqnarray}
for the Lorenz ordering of the $\kappa$GG distribution.

\section*{Acknowledgements}
A. Vallejos would like to thank Dr. Ivan Voitalov from the Department of Physics at the Northeastern University in Boston, for valuable suggestions.
% \section*{References}

\bibliography{mybibfile}

\end{document}